\pdfoutput=1

\documentclass[
    ,final            
  ]
  {aipproc}

\layoutstyle{8x11double}

\usepackage{amssymb,amsmath}

\begin{document}

\title{Universal Aspects of Deconfinement in 2+1 Dimensions}



\author{Lorenz von Smekal, Sam R.~Edwards and Nils Strodthoff}{
 address={
 Institut f\"ur Kernphysik, 
 Technische Universit\"at Darmstadt,
 Schlossgartenstra{\ss}e 9, 64289 Darmstadt}
}

\begin{abstract}
The 2+1 dimensional pure $SU(N)$ gauge theories with $N\le 4$ 
are candidates for applying the powerful tools of scaling and
universality to their  deconfinement transitions at finite temperature.   
The corresponding 2 dimensional $q$-state Potts models with $q\le 4$
have $2^\mathrm{nd}$ order transitions, and we can exploit many exact results 
to obtain accurate critical couplings, transition temperatures, critical
exponents, and the leading behavior of the continuum string tension
near the phase transition on one side, together with its dual on the other. 
Thereby, the self-duality of the 2$d$ spin models is reflected in a
duality between spacelike vortices and confining electric fluxes. 
We also discuss the relevance of center symmetry and the corresponding
vortices for confinement in full QCD when the electromagnetic
interactions of fractionally charged quarks are included.  
\end{abstract}
\maketitle

\subsection{Introduction}

\vspace{-.2cm}
The finite temperature deconfinement transition in $SU(N)$ gauge
theories in $d+1$ dimensions is very well understood in terms of the
spontaneous breakdown of their global $Z_N$ center symmetry
\cite{Greensite:2003bk}.  
Static fundamental charges represented by Polyakov loops $P(\vec x)$ transform 
under this symmetry like spins $s_i$ in a $d$ dimensional $q$-state Potts
model with $q = N$ and Hamiltonian \cite{Wu:1982ra},      
\begin{equation} 
\mathcal H = - J\sum_{\langle i,j\rangle} \delta_{s_i,s_j} - H\sum_i
\delta_{s_i,0} \;, \;\; s_i = 0,1,\dots q-1\;,
\label{PottsH}
\end{equation}

\vspace{-.2cm}
\noindent
with nearest neighbor coupling $J$. A non-zero external field $H$,
inversely related to the quark mass $m_q$, may be included to mimic the leading
effect of heavy dynamical quarks. When $1/m_q =0$, the Polyakov loop
develops a non-zero expectation value only in the deconfined, $Z_N$-broken
phase, while the expectation value of $P(\vec x) $ vanishes in the
disordered, confined phase much like the spontaneous magnetization in
the spin model. 

This is well described in terms of spacelike center
vortices which play the role of spin interfaces. They separate regions
where the Polyakov loop differs by a phase $z\in Z_N$, so that their
proliferation disorders the Polyakov loop and leads to confinement.
The suppression of spatial center vortices at high temperatures 
coincides with the ordering of the Polyakov loop, and
their free energy offers an elegant order parameter for the transition 
\cite{deForcrand:2001dp}.   

When the phase transition is of second order its description is
universal \cite{Svetitsky:1982gs}. In 3+1 dimensions this only applies
to $SU(2)$, where spatial center vortex sheets show the universal
behavior of interfaces in the 3$d$ Ising model
\cite{deForcrand:2001nd}. In 2+1 dimensions, on the other 
hand, both $SU(2)$ and $SU(3)$ exhibit a second-order
deconfinement transition, and many exact results from the
interfaces in the 2$d$ Ising and $3$-state Potts models can be
fruitfully exploited for precision studies of the (2+1)$d$
gauge theory \cite{Edwards:2009qw,vonSmekal:2010xz,Strodthoff:2010dz}. 
Even for $SU(4)$, which was previously found to have a weak
first-order transition \cite{Holland:2007ar}, at least
approximate Potts scaling can be observed in a wide range of
intermediate length scales near  
criticality \cite{deForcrand:2003wa,Strodthoff:2010dz}.

\vspace{-.4cm}

\subsection{Electric Fluxes and Self-Duality in 2+1d}

\vspace{-.2cm}
Interfaces in the spin models are typically introduced as
frustrations along which the coupling of adjacent spins favors
cyclically shifted spin states rather than parallel ones for the usual
ferromagnetic couplings $J>0$. They form $d-1$ dimensional surfaces dual
to links at which the $\delta_{s_i,s_j} $ in Eq.~(\ref{PottsH}) are replaced by 
$\delta_{s_i,s_j+ m \bmod q} $, and are conveniently studied by
introducing analogous cyclically shifted boundary conditions.
 
The ratios  $R_q^{(m,n)} \equiv Z_q^{(m,n)}/Z_q^{(0,0)} $ of $q$-state
Potts model partition functions $Z_q$ with combinations of cyclically shifted
($m, n$)-boundary conditions on a 2$d$ torus  over the periodic ensemble with
$m=n=0$  then define the 
interface free energies per
temperature as $F_I^{(m,n)} = - \ln   R_q^{(m,n)}$.
 
Center vortex free energies in $d$+1 dimensional $SU(N)$ are 
defined from analogous ratios of partition functions with
't Hooft's twisted boundary conditions \cite{'tHooft:1979uj}  
over the periodic ensemble. 
These are classified either as magnetic twists, defined in purely
spatial planes, or as temporal twists in the planes oriented along the
Euclidean time direction. The former are irrelevant for the phase
transition \cite{vonSmekal:2002gg} and will not be further discussed here.
The latter are labeled by $k_i=0,1,\dots N\!-\! 1$, and introduce spatial
center vortices perpendicular to the spatial direction $i$ of the
twist. These separate regions where fundamental Polyakov loops differ
by a center phase $z = e^{2\pi i \, k_i/N}$ and are thus like the spin
interfaces in the Potts models.    

Exact maps between the spin systems and their dual theories,
in terms of disorder variables on the dual lattice, are provided by 
Kramers-Wannier duality \cite{Savit:1979ny}. 
For the  $q$-state Potts models in 2 dimensions, these dual theories are
 $q$-state Potts models again, but at a dual temperature $\widetilde T$
which is swapped around criticality at $T_c$.
A simple proof for infinite lattices is given in
\cite{Wu:1982ra}.  Duality transformations in a finite volume do not
preserve boundary conditions, however. Periodic boundary conditions on
one side generally correspond to fluctuating boundary conditions on the other.
From the exact finite-volume duality transformation
for the $2d$ $q$-state Potts models 
\cite{vonSmekal:2010xz}, 
\vspace{-.2cm}
\begin{equation}
R_q^{(m,n)}(\widetilde K) = \frac{ {\textstyle \sum_{r,s} } 
\, e^{\frac{2\pi i}{q} (rn-sm)}\; R_q^{(r,s)}(K)}{ { \textstyle
  \sum_{r,s}} \; R_q^{(r,s)} (K)}  \; , \label{sd}
\end{equation}

\vspace{-.2cm}

\noindent
where the coupling per temperature  $K=J/T$ and
its dual $\widetilde K = J/\widetilde T$  are related by
$(e^{\widetilde K}-1)(e^K -1) = q$, with 
criticality at $K= \widetilde K = K_c = \ln(1+\sqrt q)$.

\begin{figure}

\parbox{\linewidth}{
\includegraphics[width=0.9\linewidth,trim=.8cm 0 10.3cm 0
0,clip=true]{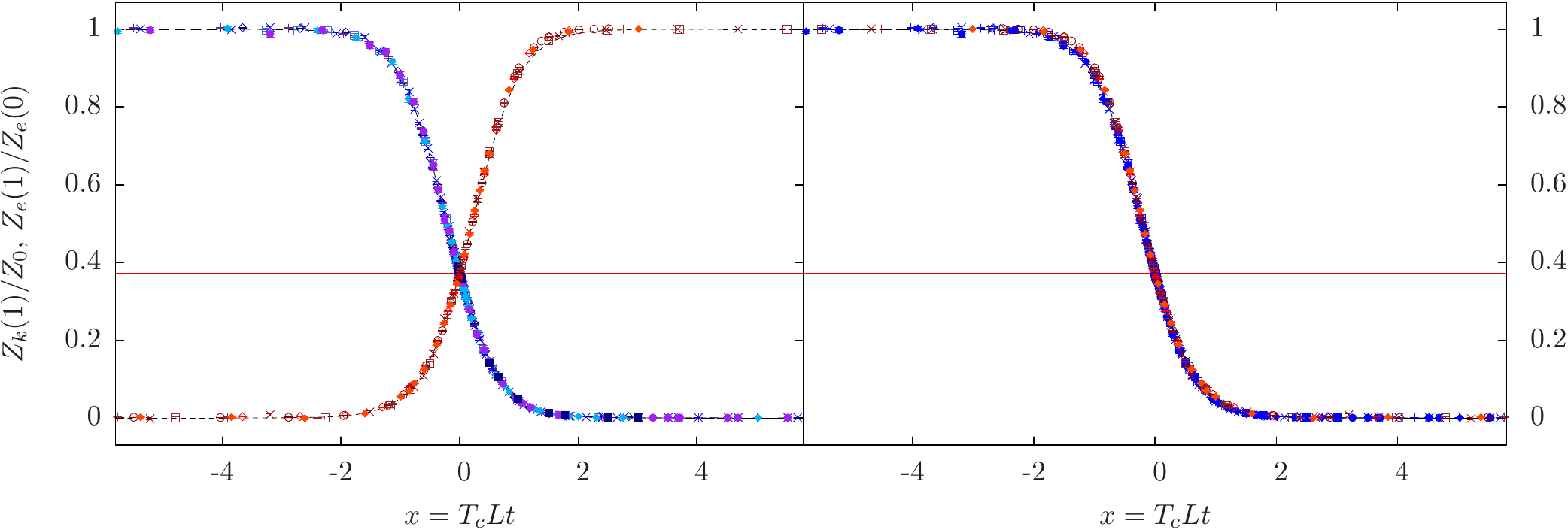}

\vspace*{-4.8cm}
{\footnotesize\hskip .8cm $R_k(x) $ \hskip 4.8cm
  $R_e(x) $ }
\vspace{4.2cm}}

\caption{The ratios of $SU(2)$ partition functions $R_k(x)$ for
  twist $(1,0)$ and $ R_e(x) $ for electric flux $(1,0)$ over the FSS
  variable $x\propto \pm L/\xi^\pm$ (data here for
  $N_t= 4$ and  $N_s$ up to $96$). 
\vspace{-.4cm}} 
\label{fig1}
\end{figure}

With $q=N$, the discrete 2$d$ $Z_N$-Fourier transform (\ref{sd}) precisely
resembles 't Hooft's relation between the ratios $R_k(\vec k) =
Z_k(\vec k)/Z_k(0)$, with temporally twisted $\vec k $ b.c.'s over the
periodic ensemble, and those of electric fluxes $\vec e$ relative to
the no-flux ensemble with fluctuating temporal twist,
\vspace{-.6cm}
\begin{equation} 
R_e(\vec{e}) = \frac{Z_e(\vec e)}{Z_e(0)}  
= \frac{{\textstyle \sum_{\vec{k}}}\; e^{2\pi\text{i}\, \vec{e}\cdot
  \vec{k} /N}\,R_k(\vec{k})}{{\textstyle \sum_{\vec{k}}} \;
R_k(\vec{k}) }\; . 
\end{equation}

\vspace{-.2cm}

\noindent
In the (2+1)$d$ gauge theory, temperature is the same on both sides of
the $Z_N$-Fourier transform. As a consequence of the self-duality of
the spin model, however, the free energies of spatial center vortices
and those of the confining electric fluxes are mirror images of one another 
within the universal scaling window around a second order phase transition.
This is shown for $SU(2)$ in Fig.~\ref{fig1}, where $\widetilde T
\leftrightarrow T$ in the spin model amounts to $x \leftrightarrow -x$, 
and the same is true for $SU(3)$ \cite{Strodthoff:2010dz}.

\begin{figure}
\parbox{\linewidth}{\hskip .5cm
\includegraphics[width=0.8\linewidth,trim=0 0 0 0
0,clip=true]{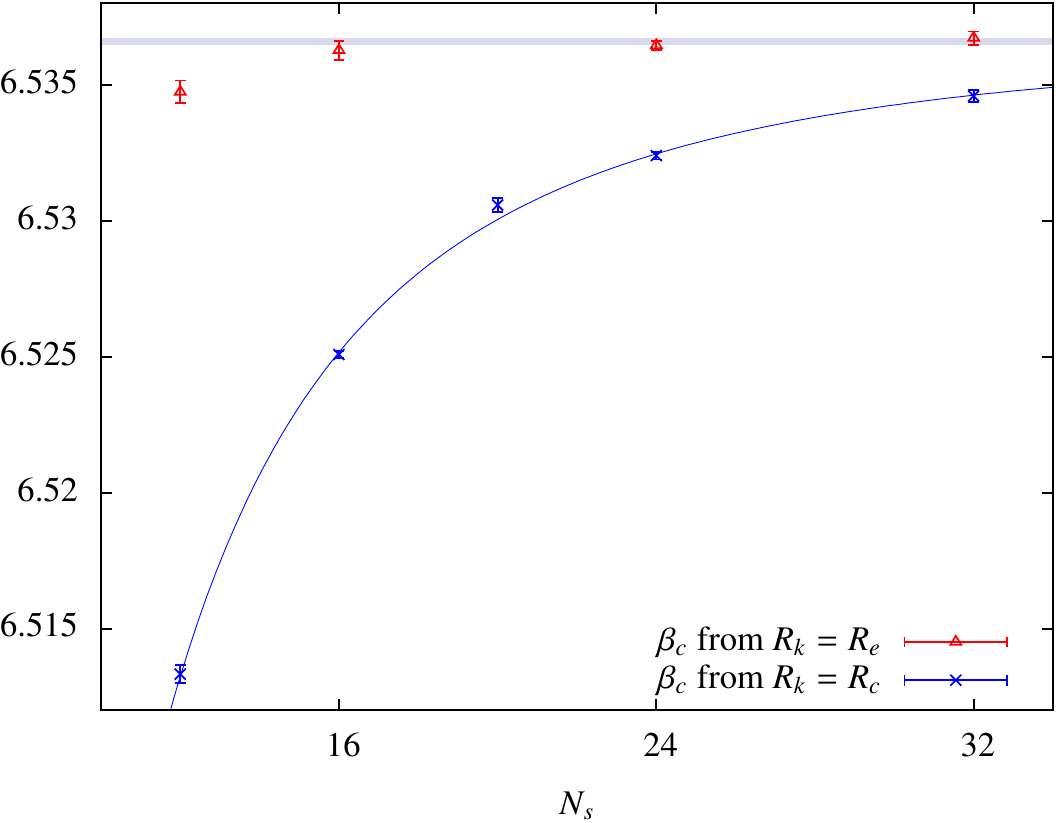}

\vspace*{-.2cm}
\caption{Critical couplings for (2+1)$d$ $SU(2)$ ($N_t= 4$) from
  self-duality compared to those of Ref.~\cite{Edwards:2009qw}, with
  $N_s$ up to 96 and the infinite-volume extrapolated result
  $\beta_c=6.53661(13) $, shown as the narrow grey band here.
\vspace{-.4cm}}} 
\label{fig3}
\end{figure}

\vspace{-.4cm}

\subsection{Critical Couplings and Finite-Size Scaling}

\vspace{-.2cm}
The by far most efficient method to determine critical couplings
$\beta_c$ from simulations in finite volumes is based on self-duality
\cite{Strodthoff:2010dz}: with that one has $R_e(\beta ) =
R_k(\beta)$ for matching $\vec e$ and $\vec k$ at lattice coupling $\beta
=\beta_c$. Defining pseudo-critical couplings in a finite volume by this
requirement,  $R_e(\beta ) = R_k(\beta)$, it is straightforward to
show that the leading finite-size corrections do not change its
value. This is because they give the same $\beta$-independent
contribution to both, $R_e$ and $R_k$, and hence move their intersection
point vertically, without any shift in $\beta$. 

In Fig.~\ref{fig3} we demonstrate how well this works by comparing 
the results from self-duality in (2+1)$d$ SU(2) to the
high-precision determination of \cite{Edwards:2009qw}.  The method used
there allowed a finite-size scaling ansatz to fit the pseudo-critical
couplings shown as the lower points in Fig.~\ref{fig3}. Their
infinite-volume extrapolation led to the value $\beta_c = 6.53661(13)$
for $N_t=4$ time slices and spatial lattice sizes up to $N_s=96$,
which is indicated together with its small error by the grey band in
the figure.  Within the present statistical errors, the 
pseudo-critical couplings from self-duality are fully consistent 
with this narrow band from the previous (far more expensive)
determination already for $N_s=16$, {\it i.e.}, we practically
reproduce the infinite volume result with an aspect ratio of only
$4:1$. The weighted mean from $ N_s=16, 24 $ and 32 yields $\beta_c =
6.53651(13)$. 

With accurate critical couplings we can assess the finite-size scaling
near the second order phase transition at $t= T/T_c-1 = 0$, where
generalized couplings such as the vortex-ensemble ratios $R_k$, for 
sufficiently large sizes $L$, only depend on $L^{1/\nu} t$ in a
universal way. This can be used, for example, to extract the
correlation-length critical exponent $\nu $ from the slope $s$ of the
center vortex free energies $F_k(\beta ) = -\ln R_k $ in $\beta $ at
$\beta_c$. Because $t\propto (\beta -\beta_c)$, the slope grows
with the spatial lattice size as $s \sim N_s^{1/\nu}$. Our present data
for (2+1)$d$ $SU(N)$ in this way yields $\nu = 0.99(2)$, $0.82(4)$
and $0.60(5)$ for $N=2,3$ and 4, as compared to $\nu = 1,\, 5/6$ and
$2/3$ for the 2$d$ Potts models with $q=2,3$ and 4, respectively.   

\begin{figure}
\parbox{\linewidth}{\hskip .3cm
\includegraphics[width=0.8\linewidth,trim=0 0 0 0
0,clip=true]{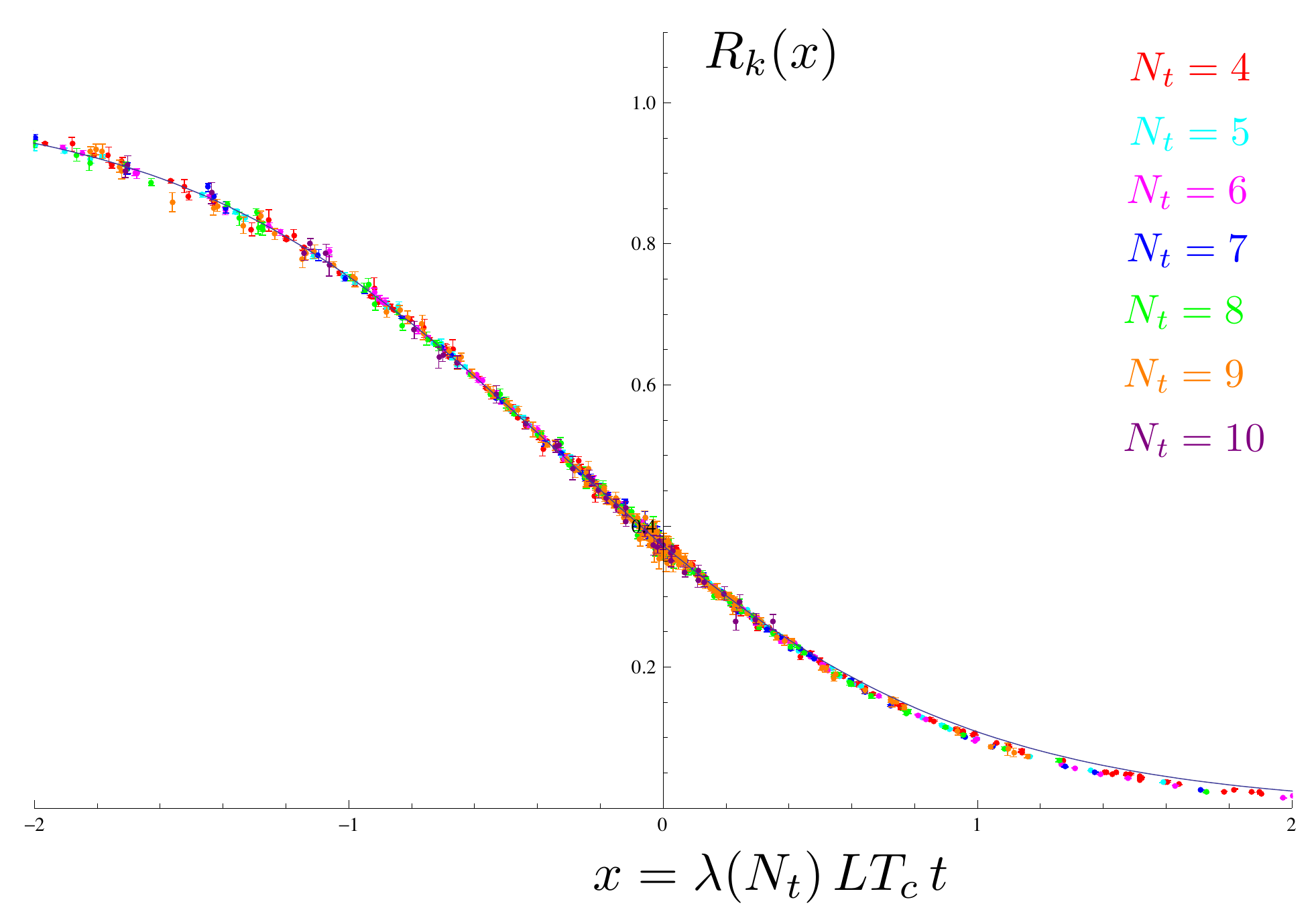}

\vspace*{-.2cm}
\caption{$SU(2)$ center-vortex partition functions with
  $N_t$-dependent rescaling of $x$ for $N_t= 4$ to 10 and
  $N_s$ up to $96$.   
\vspace{-.4cm}}} 
\label{fig2}
\end{figure}

Moreover, for $SU(2)$ this dependence of the center vortex free
energies is known exactly, from the analytically computable universal
scaling functions $F_I(x)$ for the interface free energies in the 
$2d$ Ising model \cite{vonSmekal:2010xz,1999JPhA.32.4897W}. One
obtains $F_k(x) = F_I(-\lambda x)$ where the single non-universal factor
$\lambda   $ can be accurately determined from one-parameter fits
near $x=LT_c t =0$. This was done for $N_t = 4$ to 10
in \cite{vonSmekal:2010xz}, which allows the continuum extrapolation
$\lambda =  1.354(25)$ for $N_t \to \infty$. We can furthermore
compensate discretization effects by simply rescaling the finite-size
scaling variables on different $N_t$ lattices as $x\to \lambda(N_t)
x$, to observe good (continuum) scaling around criticality for all
available $N_s$ {\em and} $N_t$, as shown in Fig.~\ref{fig2}. 

Other immediate consequences of this extrapolation are, {\it e.g.},
the  behavior of the continuum string tension and its dual around
the deconfinement transition in (2+1)$d$ $SU(2)$:
$\sigma/T_c^2  = 2.387(44) \, |t| + \cdots$, for $t\to 0^-$, and
$\tilde\sigma/T_c = 2.387(44) \, t + \cdots $, for $t\to 0^+$,
respectively \cite{vonSmekal:2010xz}.

\vspace{-.4cm}

\subsection{Conclusions and Outlook}

\vspace{-.2cm}
Universality and scaling allow one to study in detail 
the deconfinement transition in the 2+1 dimensional $SU(2)$, $SU(3) $,
and at least approximately also $SU(4)$ gauge theories, by relating
their spatial center vortices to interfaces in 2$d$ Potts models for
which many exact results are available.  The self-duality of these
models is reflected in the gauge theories: around criticality, 
the free energies of the confining electric fluxes are mirror images
of those of spatial center vortices. We demonstrated how this can be 
exploited to remove the leading finite-size corrections in the
determination of critical couplings from numerical simulations. 
These then allow detailed finite-size scaling analyses and simple
continuum extrapolations such that the consequences of universal
scaling are carried over to the continuum gauge theory.

\begin{figure}
\parbox{\linewidth}{\hskip .5cm
\includegraphics[width=0.8\linewidth,trim=0 0 0 0
0,clip=true]{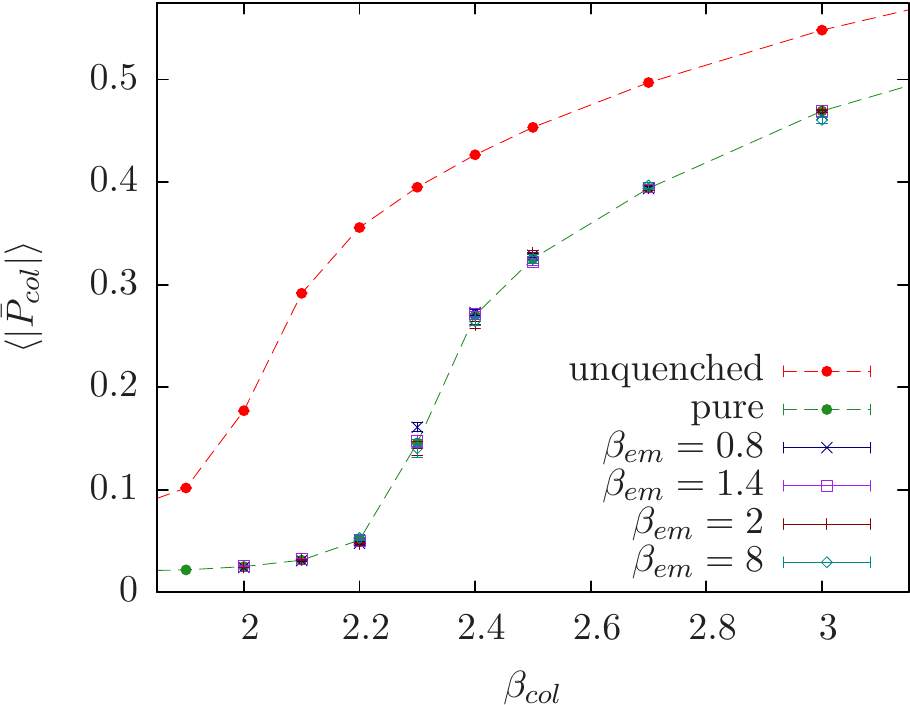}

\vspace*{-.2cm}
\caption{$SU(2)$ Polyakov loop with half-integer
  charged dynamical quarks ($\kappa =
  0.15$) on an $8^3\times 4 $ lattice with hot $U(1)$ start to provide
  $Z_2$ disorder in Coulomb phase (at large
  $\beta_{em}$).  
\vspace{-.4cm}}} 
\label{fig4}
\end{figure}

We close with a short remark on the relevance of center symmetry and center
vortices for confinement in full QCD when including the electromagnetic
interactions of fractionally charged quarks. Dynamical quarks
explicitly break center symmetry in the same way as the external
field $H$ in (\ref{PottsH}).
When electromagnetic $U(1)$ couplings are
included, however, this may become more like a fluctuating external
field. Because of the quarks' fractional charges, the $U(1)$ gauge
action is then unable to remove initial disorder as it evolves from
strong to weak coupling for integer electric charges. This is
demonstrated for a $SU(2) \times U(1)/Z_2 $ toy model with
half-integer charged quarks in Fig.~\ref{fig4}, where the quenched
behavior is recovered for  the $SU(2)$ Polyakov loop with a hot $U(1)$
start, far into the Coulomb phase for integer charges, see
\cite{Edwards:2010ew}.


\vspace{.1cm}

\noindent\textbf{Acknowledgements:} This work was supported by the
Helmholtz International Center for FAIR within the LOEWE program of
the State of Hesse, the Helmholtz Association, Grant VH-NG-332, and the
European Commission, FP7-PEOPLE-2009-RG No.~249203. Simulations were
performed on the high-performance computing facilities of eResearch
SA, South Australia. 


\vspace*{-.4cm}

\bibliographystyle{aipproc}   

\begin{thebibliography}{9}

\vspace*{-.2cm}

\bibitem{Greensite:2003bk}
J.~Greensite, Prog.\ Part.\ Nucl.\ Phys.\ {\bf 51} (2003) 1.

\bibitem{Wu:1982ra}
  F.~Y.~Wu,
  Rev.\ Mod.\ Phys.\  {\bf 54} (1982) 235.

\bibitem{deForcrand:2001dp}
  Ph.~de Forcrand and L.~von Smekal,
  Nucl.\ Phys.\ Proc.\ Suppl.\  {\bf 106} (2002) 619.

\bibitem{Svetitsky:1982gs}
  B.~Svetitsky, L.~G.~Yaffe,
  Nucl.\ Phys.\  B {\bf 210} (1982) 423.


\bibitem{deForcrand:2001nd}
  Ph.~de~Forcrand and L.~von Smekal, Phys.\ Rev.\ D {\bf 66} (2002) 011504(R).

\bibitem{Edwards:2009qw}
  S.~Edwards, L.~von Smekal,
  Phys.\ Lett.\  B {\bf 681} (2009) 484.

\bibitem{vonSmekal:2010xz}
  L.~von Smekal, S.~R.~Edwards and N.~Strodthoff,
  arXiv:1012.0408 [hep-lat].

\bibitem{Strodthoff:2010dz}
  N.~Strodthoff, S.~R.~Edwards and L.~von Smekal,
  arXiv:1012.0723 [hep-lat].

\bibitem{Holland:2007ar}
  K.~Holland, M.~Pepe, U.~J.~Wiese,
  JHEP {\bf 0802} (2008) 041.

\bibitem{deForcrand:2003wa}
  Ph.~de Forcrand and O.~Jahn,
  Nucl.\ Phys.\ Proc.\ Suppl.\  {\bf 129} (2004) 709.

\bibitem{'tHooft:1979uj}
  G.~'t~Hooft, 
  Nucl.\ Phys.\ B {\bf 153} (1979) 141.

\bibitem{vonSmekal:2002gg}
  L.~von Smekal and Ph.~de Forcrand,
  Nucl.\ Phys.\ Proc.\ Suppl.\  {\bf 119} (2003) 655.


\bibitem{Savit:1979ny}
  R.~Savit,
  Rev.\ Mod.\ Phys.\  {\bf 52} (1980) 453.

\bibitem{1999JPhA.32.4897W}
  M.-C.~Wu, M.-C.~Huang, Y.-P.~Luo and T.-M.~Liaw, 
  J.\ Phys.\ A:\ Math.\ Gen.\ {\bf 32} (1999) 4897.

\bibitem{Edwards:2010ew}
  S.~R.~Edwards, A.~Sternbeck and L.~von Smekal,
  arXiv:1012.0768 [hep-lat].

\end{thebibliography}

\end{document}